\documentclass{IEEEtran4PSCC}

 \usepackage[subtle,tracking=normal ]{savetrees}

   \usepackage[dvips]{graphicx}

\usepackage{amsfonts, amssymb, amsthm}

\newtheorem{remark}{Remark}
\usepackage{booktabs}
\usepackage[table,xcdraw]{xcolor}
\usepackage[cmex10]{amsmath}
\usepackage[noadjust]{cite}

\makeatletter
\let\old@ps@headings\ps@headings
\let\old@ps@IEEEtitlepagestyle\ps@IEEEtitlepagestyle
\def\psccfooter#1{%
    \def\ps@headings{%
        \old@ps@headings%
        \def\@oddfoot{\strut\hfill#1\hfill\strut}%
        \def\@evenfoot{\strut\hfill#1\hfill\strut}%
    }%
    \def\ps@IEEEtitlepagestyle{%
        \old@ps@IEEEtitlepagestyle%
        \def\@oddfoot{\strut\hfill#1\hfill\strut}%
        \def\@evenfoot{\strut\hfill#1\hfill\strut}%
    }%
    \ps@headings%
}
\makeatother

\psccfooter{%
        \parbox{\textwidth}{\hrulefill \\ \small{23rd Power Systems Computation Conference} \hfill \begin{minipage}{0.2\textwidth}\centering \vspace*{4pt} \includegraphics[scale=0.06]{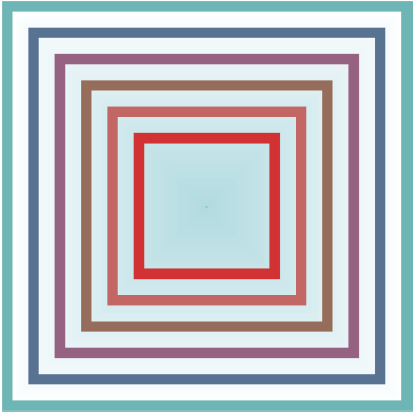}\\\small{PSCC 2024} \end{minipage} \hfill \small{Paris, France --- June 4 -- 7, 2024}}%
}

\begin{document}
%
\title{Optimization-based Framework for Selecting Under-frequency Load Shedding Parameters}

\author{
\IEEEauthorblockN{Waheed Owonikoko, Mazen Elsaadany, Amritanshu Pandey, 
Mads R.Almassalkhi}
\IEEEauthorblockA{Department of Electrical Engineering \\
University of Vermont\\
Burlington, Vermont}
}


\maketitle


\begin{abstract}
High penetration of renewable resources results in a power system with lower inertia and higher frequency sensitivity to power imbalances. Such systems are becoming increasingly susceptible to frequency collapse during extreme disturbances. Under-Frequency Load Shedding (UFLS) is a last-resort protection scheme and acts as an emergency brake by shedding load to arrest frequency decline. Current and emerging efforts to optimize UFLS settings and frequency thresholds are mostly network agnostic, ignoring network spatial information. With the prevalence of Distributed Energy Resources (DERs) in the high-renewable paradigm, the power grid is becoming more bidirectional, making some locations in the network less effective for UFLS action than others. This work proposes a Mixed Integer Linear Program that optimizes the UFLS setpoints (prioritizing one location over another) to minimize frequency deviation and load-shed for a given disturbance. The formulation considers system information and DER generation mix at different network locations, increasing model fidelity. The formulation also captures the discrete nature and practical time delays and deadbands associated with UFLS using a minimal set of binary variables, reducing problem complexity. We empirically validate the optimization approach on the dynamic IEEE 39-bus system for performance metrics, including frequency nadir, steady-state frequency and total load shed.
\end{abstract}




\thanksto{\noindent This work was supported by DOE/PNNL via award DE-AC05-76RL01830.}

%
%








\section{Introduction}
Renewable energy resources (RES) are being deployed at an increasing rate. This has resulted in lower system inertia and larger frequency sensitivity to power imbalances. As such, the system is more prone to under-frequency events during disturbances like the loss of a major generator, which can lead to a blackout if not dealt with promptly. To arrest the frequency decline before it reaches undesirable levels, appropriate control measures must be established. Under-frequency load shedding (UFLS) is the power system's emergency control mechanism~\cite{kundur2022power}, designed to arrest frequency decline during large/extreme under-frequency events by shedding load and easing imbalances between generation and demand.
 

UFLS can be divided into two main categories: i) conventional UFLS, which entails shedding a certain (pre-defined) amount of load at specified frequency thresholds, ii) adaptive UFLS schemes, which shed load based on the rate of change of frequency (RoCoF) and/or the frequency deviation from nominal ~\cite{liu2014optimal, ghaderi2017dynamic,sun2021underfrequency,7337458}. Note, that adaptive UFLS schemes can further be classified into several sub-categories.



Conventional UFLS schemes operate under outdated assumptions, which may lead to sub-optimal load-shedding~\cite{adiyabazar2020optimal, he2019decentralized, 7337458, sigrist2012performance}. Many methods of optimizing the conventional UFLS scheme have been proposed in recent literature. The authors in \cite{rafinia2020stochastic, ceja2012under} and \cite{amraee2017probabilistic} present a Mixed Integer Linear Program (MILP) formulation to optimally obtain frequency thresholds and load shedding amount for each stage. The work optimizes the setpoints in a conventional UFLS setting, considering time delay, system inertia, damping, and uncertainty from solar PV generation. However, the MILP formulations presented reduce the network and use the center of inertia (CoI) to capture system dynamics losing network spatial information in the process. With the increased penetration of Distributed Energy Resources (DERs) in the grid, the power system is subject to bidirectional injections, with some areas having distributed generation that exceeds demand (back-feeding). Triggering a UFLS relay, at a back-feeding substation would lead to a loss of (net) generation and will worsen frequency conditions. Therefore, considering network spatial information and information about DER generation in the UFLS scheme design is crucial as the grid transitions into a more renewable and bidirectional future.


On the other hand, adaptive UFLS methods may also shed load sub-optimally, as local RoCoF measurements alone can be inaccurate are insufficient due to noise. While the CoI RoCoF is used in adaptive schemes~\cite{milano2017rotor}, precise CoI RoCoF measurements require PMUs at all generator buses, increasing cost and adding a significant communication overhead~\cite{RoCoF_Bad}. Several adaptive UFLS schemes have also been proposed. The CoI RoCoF is approximated in~\cite{sun2021underfrequency} from local measurements based on the inflection points of the local frequency.  In~\cite{zuo2019impact}, a RoCoF-based local UFLS system was proposed; this method estimates RoCoF and frequency using PMU measurements which are then used to determine the amount of load to be shed. In~\cite{terzija2006adaptive}, the author presents a nonrecursive Newton-type approach for estimating frequency and RoCoF which are then used to determine the load shed. However, RoCoF is sensitive to noise, which can cause over/under load shedding.

Adaptive UFLS schemes come with several practical implementation challenges~\cite{RoCoF_Bad}, and are still agnostic to network spatial information and DER generation information~\cite{sun2021underfrequency,zuo2019impact,terzija2006adaptive}. Static conventional UFLS schemes operate on outdated assumptions leading to sub-optimal performance. While MILP-based formulations have been used to optimize conventional UFLS scheme design \cite{amraee2017probabilistic, rafinia2020stochastic}, they have been network agnostic and questions of where to shed the load, how much load should be shed, and whether some buses are more suitable for UFLS than others are still unanswered. 
Previous work presented in \cite{Dofler,poolla2018virtual} consider the network spatial information for continuously controllable resources such as inverters providing virtual inertia/damping. However, UFLS schemes do not entail continuously controllable resources since load shedding is discrete in nature.

 Hence, this work proposes an optimization framework wherein UFLS setpoints are adapted regularly by considering changing grid conditions, network spatial information and DER generation information to UFLS scheme-design. The paper's contributions are summarized as follows:


\begin{enumerate}
    \item  \textbf{Improved model fidelity:} We design a high-fidelity UFLS scheme with a MILP-based optimization that incorporates spatial information to optimize UFLS parameters. The formulation also incorporates temporal information about DER generation at various bus locations to mitigate the adverse effects of triggering a UFLS relay at locations where back-feeding occurs. Furthermore, DER generation information is also used to maximize the performance of the UFLS scheme while minimizing the amount of load shed.
    

    \item  \textbf{Reduced solution complexity:} We reduce complexity in MILP formulation by using a single binary variable to a) indicate when the frequency is below the threshold and b) when the load-shedding action is activated.
    


    \item \textbf{Empirical validation:} We demonstrate the efficacy and benefits of our proposed approach with simulation-based analysis on validated dynamic test cases.
    \end{enumerate}

The rest of the paper is arranged as follows: Section~\ref{Sec_SystemModel} describes the dynamic model used in the UFLS scheme design and validates the linear model against present non-linear dynamic models. Section \ref{Sec_OptFormulation} describes the optimization problem formulation. Section~\ref{Sec_Methodology} covers the UFLS parameter selection methodology.  Section~\ref{Sec_ResultsDiscussion } and Section~\ref{Sec_conclusion } discuss the results and conclusion, respectively.

\section{System Modeling and Validation }\label{Sec_SystemModel}
\subsection{System Modeling}
The bulk power system dynamics are governed by AC power flow equations, swing equations, generator governor and turbine dynamics, automatic voltage regulators, and load models. For the analysis in this paper, we adopt a simplified dynamics model of the power grid that considers the generator governor, swing equations, and DC power flow equations. In Section~\ref{Sec_Model_Validation}, we evaluate the efficacy of this simplified model against the nonlinear model. The simplified dynamic network model is adapted from~\cite{poolla2018virtual} and~\cite{9326147}. In the model, we denote deviations in voltage phase angle  and  angular frequency from nominal, at each bus $n\in \mathcal{B} :=\{1,\hdots, N\}$, as $\theta_n$ and $\omega_{n}$, respectively. 
That is, at nominal steady state, $\theta_n=0 $ rad and  $\omega_{n}=0$ pu.

Let a power system be described by a weighted graph $\mathcal{G}=(\mathcal{B}, \mathcal{E})$. $\mathcal{B}$ is the set of buses is composed of the union of disjoint sets of generator ($\mathcal{B}_G$) and non-generator ($\mathcal{B}_L$) buses, i.e. $\mathcal{B} = \mathcal{B}_G \cup \mathcal{B}_L$ and $\mathcal{B}_G \cap \mathcal{B}_L = \emptyset$. $\mathcal{E}$ denotes the transmission lines connecting the buses in the network, each line is weighted with its corresponding line susceptance. Let the number of buses in the system be $N$ and the number of edges (transmission lines) be $N_e$. The bus-line incidence matrix, $C \in \mathbb{R}^{N \times N_e}$, is
\begin{align}
C_{je}=
    \begin{cases}
        1 & \text{if node } j \text{ is the source of edge } e\\
        -1 & \text{if node } j \text{ is the sink of edge } e\\
        0 & \text{otherwise } 
    \end{cases}.
\end{align}

Denote $B_e \in \mathbb{R}^{N_e \times N_e}$ the diagonal matrix with entries $B_{ij}$. The weighted graph Laplacian matrix, $L \in \mathbb{S}^n$ where $\mathbb{S}^n$ is the set of symmetric $n \times n$ matrices, can be defined as
\begin{align}
    L=CB_eC^T.
\end{align}

Thus, let $\theta_G$ and $\omega_G$ denote vectors with the angles and frequencies of generator buses while $\theta_L$ and $\omega_L$ denote vectors of angles and frequencies at non-generator buses. Collating these vectors yields the full network's bus angles and frequencies
\begin{align}\label{eqn:Theta-OmegaDef}
{\theta} := \begin{bmatrix}
    \theta_G \\
    \theta_L
\end{bmatrix} \hspace{5pt} 
{\omega} := \begin{bmatrix}
    \omega_G \\
    \omega_L
\end{bmatrix}
\end{align}
To incorporate the dynamics of generator bus angles and frequencies, the swing equations are used for every generator bus $n\in \mathcal{B}_G $ as
\begin{align}
\dot {\theta}_{G,n} &= \omega_\text{base}  \omega_{G,n} \label{eqn:ThetaG_eqn}\\
M_n \dot {\omega}_{G,n} &= P_{Line,n} -D_n{\omega}_{G,n} + P_{m,n} + P_{\text{UFLS},n} + \Delta_n, \label{eqn:OmegaG_eqn}
\end{align}
where $M$ and $D$ are the generator inertia and damping coefficients, respectfully, $P_{Line,n}$ is the deviation in the net power injection to bus $n$ from nominal, and $P_{m,n}$ is the deviation in generator mechanical power input from nominal and driven by the governor. The base frequency is denoted $\omega_\text{base}$ and is in rad/sec. Note, a factor of $\omega_\text{base}$ is needed in \eqref{eqn:ThetaG_eqn} since the frequency is in per unit and the bus angle is in radians. $\Delta_n$ denotes the disturbance occurring at bus $n$.  The net load shed at bus~$n$ is denoted $P_{\text{UFLS},n}$. If a bus does not have load or is not participating in the UFLS scheme, then the corresponding $P_{\text{UFLS},n}$ is omitted. 

\begin{remark}[Load Shed vs Power Disruption]
UFLS has historically been implemented at distribution substation, but With the prevalence of DERs, including PV generation, at the load side, distribution substations are increasingly becoming bidirectional at the T\&D interface. Thus, tripping a UFLS relay on a back-feeding substation results in a net increase in generation. With DER generation in mind, it is important to consider that UFLS schemes will not just be limited to shedding the load, but more broadly will be causing \textit{power disruptions (i.e. result in sudden drop of either net generation or load)}. The goal of any intelligent UFLS scheme is then to minimize power disruptions from back-feeding substations while shedding enough load elsewhere to arrest frequency declines. 
\end{remark}

The mechanical power input to the generator $n$ from the governor is modeled as a first-order low-pass filter with the transfer function shown below:
\begin{align}\label{eqn:Governor_TF}
P_{m,n}(s) = - \frac{K_{\text{gov},n}}{T_{\text{gov},n} s + 1} \omega_n,
\end{align}
where $K_{\text{gov},n}$ is the governor gain and $T_{\text{gov},n}$ is the governor time constant. In time domain, the mechanical power input, $P_{m,n}(t)=  -{K_{\text{gov},n}}{x_{\text{gov},n}(t)}$, yields governor dynamics:
\begin{align}
\Rightarrow \dot{x}_{\text{gov},n}=\frac{1}{T_{\text{gov},n}}(\omega_{G,n} - x_{\text{gov},n}) \qquad \forall n\in \mathcal{B}_G \label{eqn:Governor_states}.
\end{align}
The change in net power flowing into bus~$i$ is a function of the change in sum of line flows:
\begin{equation}\label{eqn:Pline_eqn}
  P_i,_{Line}= - \sum_{j\in \mathcal{N}_i} P_{ij}, \qquad \forall i \in \mathcal{B}
\end{equation}
where $P_{ij}$ is the line flow from bus $i$ to bus $j$ and $\mathcal{N}_i \subset \mathcal{B}$ is the subset of buses that are neighbors to bus~$i$. The DC-power flow is chosen to model the line flows:
\begin{equation}\label{eqn:DCPowerFlow}
P_{ij} = B_{ij} ( \theta_i - \theta_j) \qquad \forall \{i,j\} \in \mathcal{E}.
\end{equation}

\noindent From~\eqref{eqn:DCPowerFlow}, the net line flow injection at any bus $n \in \mathcal{B}$ is defined by the $n^{\text{th}}$ element of the product of the weighted graph Laplacian ($L$) and the vector $\theta$:
\begin{equation}\label{eqn:1}
  P_{Line,n}= -[L\theta]_n.
\end{equation}
For buses without a generator, the bus angles and frequencies are governed by the following:
\begin{equation}\label{eqn:theta_load_eqns}
\Pi_L L \theta = \Pi_L P_\text{UFLS} + \Pi_L \Delta,
\end{equation}
where $\Pi_L$ is the non-generator bus selection matrix satisfying $\theta_L=\Pi_L \theta$.
Differentiating~\eqref{eqn:theta_load_eqns}, the frequency dynamics at non-generating buses are given by:
\begin{equation}\label{eqn:theta_load_diff}
\Pi_L L\dot{\theta}=\omega_\text{base}\Pi_L L \omega = \Pi_L\dot{P}_\text{UFLS} + \Pi_L \dot{\Delta}.
\end{equation}
 In UFLS applications, ${P}_\text{UFLS}$ and $\Delta$ are step inputs to the system model equations, which implies that their derivatives are impulse functions that are non-zero for only an infinitesimal time duration. This simplifies the non-generator equations to the following expression~\cite{9326147}:
\begin{equation}\label{eqn:omega_Load_eqn}
\omega_\text{base}\Pi_L L \omega = 0.
\end{equation}

\noindent The bus frequencies can be expressed as a set of Differential Algebraic Equations (DAEs), where \eqref{eqn:ThetaG_eqn}-\eqref{eqn:OmegaG_eqn} govern $\theta_G$ and $\omega_G$ and~\eqref{eqn:theta_load_eqns} and \eqref{eqn:omega_Load_eqn} govern $\theta_L$ and $\omega_L$.



\noindent The set of differential equations defined in ~\eqref{eqn:ThetaG_eqn}-\eqref{eqn:OmegaG_eqn} are then discretized in time using trapezoidal integration, which yields the difference equation 
\begin{multline}\label{eqn:DiscretizedEqn}
  x[t_{n+1}] \approx x[t_n]+ \frac{\Delta t}{2}\left( Ax[t_{n+1}]   + B P_\text{UFLS}[t_{n+1}] \right. \\
    \left. + G \Delta[t_{n+1}] + Ax[t_n] + BP_\text{UFLS}[t_n] + G\Delta[t_n] \right),
\end{multline}
 \noindent where vector $x^\top:=\text{col}\{\theta^\top, \omega^\top, x_\text{gov}^\top\}$. For details on the $A,B$, and $G$ matrices, please see Appendix. 

\subsection{Model Validation} \label{Sec_Model_Validation}

\begin{figure}[t]
   \centering
	\includegraphics[trim={0cm 0cm 0cm 0cm}, clip,width=0.9\linewidth]{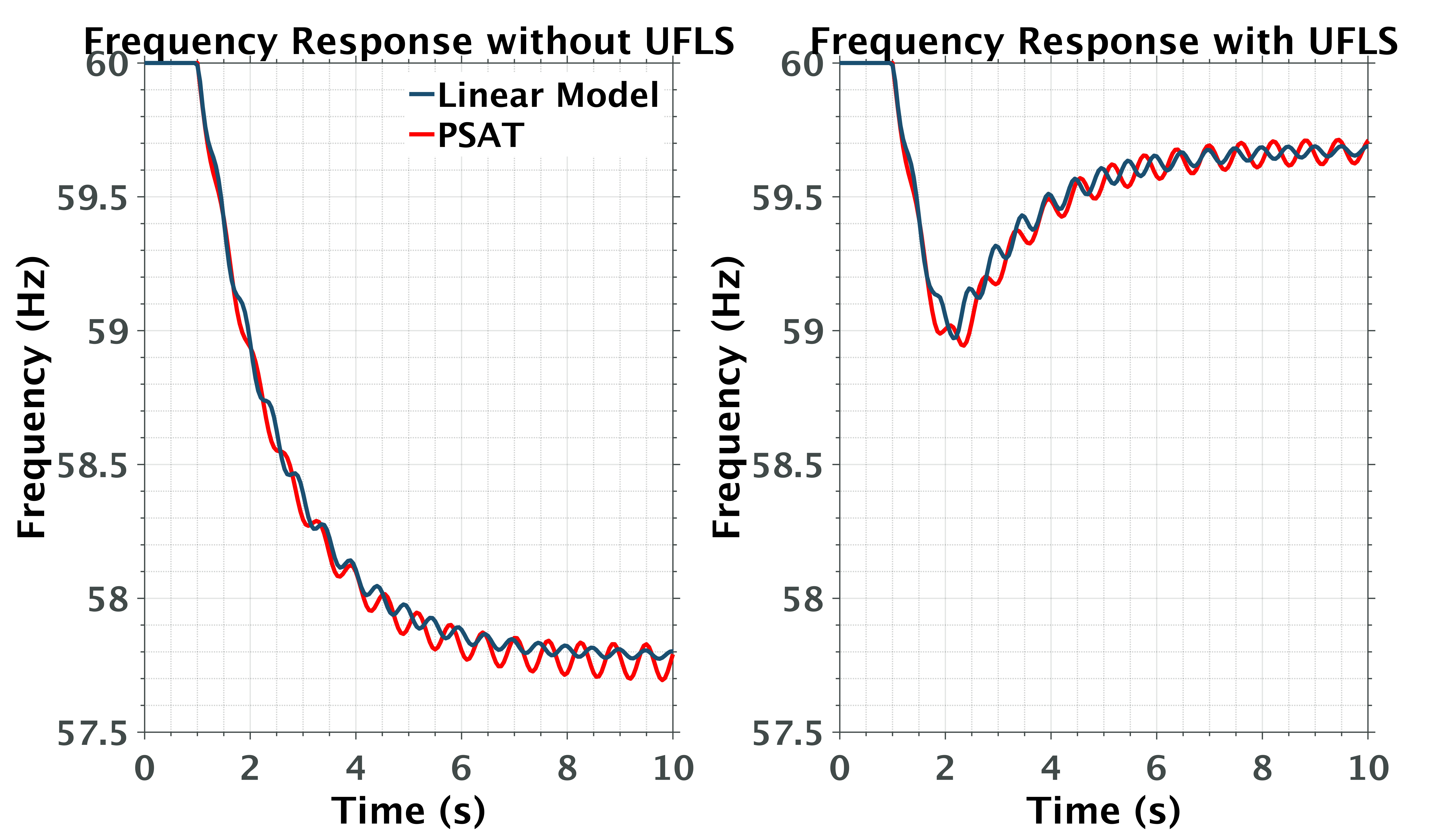}
 	\caption{Illustrative comparison of linear and nonlinear (PSAT) models with and without UFLS enabled.}
 	\label{fig:Validation}       
 \end{figure}

The linear DAE model presented in the previous section is validated against the full nonlinear dynamics model using the WECC 9~bus system. The validation is conducted via the Power System Analysis Toolbox (PSAT) \cite{PSAT_Ref}, which considers the nonlinear AC power flow equations, swing equations,  governor dynamics, automatic voltage regulator, higher order generator model, and converted constant power load models. The WECC 9~bus system parameters were obtained from PSAT. In determining how well the linear model captures the dominant nonlinear PSAT modes, the focus of validation is on the bus frequencies. 

Two scenarios are set up for validation, each of which consider a loss of generator \#3 as the disturbance, 1) without UFLS action and 2) with UFLS. The conventional UFLS frequency thresholds and load shed amounts were implemented as laid out by NERC PRC-006.  Fig~\ref{fig:Validation}, shows the post-disturbance frequency of Bus~2 obtained from both the linear model and PSAT. The maximum and average absolute error between the linear model and PSAT bus frequencies (across all buses) are calculated and summarized in Table~\ref{tab:Model_Validation_Error}. It is clear for Fig~\ref{fig:Validation} that Bus~2 frequency response is sufficiently accurate with the simplified model for the purpose of the UFLS scheme design in this paper.
\begin{table}[h]
\centering
\caption{Comparing linear system model vs PSAT's nonlinear model}
\label{tab:Model_Validation_Error}
\begin{tabular}{@{}lcc@{}}
\toprule
 & \textbf{Without UFLS} & \textbf{With UFLS} \\ \midrule
Mean Absolute Error (Hz) & 0.03 & 0.02 \\
Max Absolute Error (Hz) & 0.13 & 0.18 \\ \bottomrule
\end{tabular}
\end{table}

The linear model mimics the overall frequency trend well from the nonlinear dynamics model for both scenarios. The discrepancies in the bus frequency are largely due to the linear model employing the DC power flow and the nonlinear PSAT using the full AC power flow.

\section{Optimization Formulation}\label{Sec_OptFormulation}

The linear model discussed in Section \ref{Sec_SystemModel} is used to optimize the UFLS settings within existing guidelines laid out in NERC PRC-006. A UFLS scheme is defined by setpoints that specify the amount of load to be shed at each load bus and the corresponding frequency thresholds at which the load shedding should be triggered. A MILP problem formulation is developed to find the optimal frequency thresholds, amount of load shed and location of load shed for a given disturbance $\Delta(t)$. We extend the analysis across a set of disturbances. Three performance metrics of a UFLS scheme are considered, i) the minimum bus frequency (frequency nadir), ii) the total load shed, and iii) the steady state frequency deviation. A multi-objective objective function is formulated incorporating the three metrics. The following objective function is used:
\begin{align}\label{eqn:Objective}
\min \hspace{5pt}\gamma_1\sum_{n=1}^{N} ||\omega^n ||_{\infty} + \gamma_{2}\sum_{n\in \mathcal{B}_L}P^n_{\text{sh}}[K] + \gamma_3\sum_{n=1}^{N} |\omega^n[K]|.
\end{align}

The first term in the objective function is the sum of the maximum frequency deviation, where $\omega^n$ is a vector whose elements are the frequency deviation at bus $n$ for all timesteps considered in the time horizon. The second term is the sum of the load shed at the end of the time horizon. The third term is the sum of the steady state frequency deviation from 60 Hz for each bus. Note, if a sufficiently large time horizon will allow the bus frequencies to settle after a transient and the terminal frequencies $\omega[K]$ will be the steady state frequencies where $K$ is the length of the time horizon considered.  The parameters $\gamma_1$, $\gamma_2$ and $\gamma_3$ are the relative weights are used to prioritize the different objectives in the multi-objective objective function. In the objective function considered the cost associated with shedding load at all buses is the same therefore a common weight ($\gamma_2$) is considered for all the buses. However, if that is not the case then a location specific cost ($\gamma_2^n$) can be considered. 


\subsection{UFLS Practical System Implications}
In a UFLS scheme, the load is shed in discrete amounts and only after the bus frequency falls below a specific threshold frequency. A UFLS scheme usually includes multiple stages, each with an associated amount of load shed and a corresponding frequency threshold. Also, for the timescale of the disturbance, the shed load is not reconnected back into the system. This means that the amount of load shed should be non-decreasing over time. Furthermore, to account for noise in frequency measurements, the load is not shed based on instantaneous frequency values; instead, a deadband is added to ensure that the load shedding only occurs after being below a frequency threshold for a given time (300 ms). In addition, UFLS relays cannot instantaneously trigger a circuit breaker, and a delay between relay tripping and circuit breaker actuation needs to be considered (100ms). \cite{amraee2017probabilistic} and \cite{rafinia2020stochastic} consider these practical constraints by introducing two binary variables. One of the binary variables indicates whether the frequency is below the threshold, and the second models the deadband and only becomes 1 if the frequency is below the threshold for more than the deadband time. Extending these formulations to include the network information increases the number of binary variables by a factor of $N$, where $N$ is the number of buses. This paper ensures the UFLS practical constraints are met with half as many binary variables, reducing computational complexity. Moreover, the underlying problem formulation is a computationally tractable MILP problem compared to the non-convex formulation in \cite{amraee2017probabilistic} and \cite{rafinia2020stochastic}. 
\subsubsection{Frequency Thresholds}
The MILP formulation needs to allow the load to be shed only when the frequency at the bus is below a frequency threshold, $\omega_{sh}^i$, where the superscript $i$ denotes the stage of the UFLS scheme. The following constraints defined in \eqref{eqn:Freq_Thres} indicate whether the corresponding bus frequency is below the frequency threshold. 
\begin{subequations}\label{eqn:Freq_Thres}
    \begin{align}
        \omega^{i}_\text{sh}-\omega_n[k]-\sum_{l=1}^{k-1}\alpha_n^i[l] & \le \alpha_n^i[k] \le  1+\omega^{i}_\text{sh}-\omega_n[k] \label{eqn:opt_thresh}\\
        \alpha_n^i[1] &=0  \quad \land \quad 
        \alpha_n^i[k] \quad \in \{0,1\} 
        \label{eqn:opt_binInit} 
    \end{align}
\end{subequations}
for $i=1,\hdots,N_\text{{UFLS}}$. $k=1,\hdots,K$, $n=1,\hdots,N_{L}$, where $N_\text{{UFLS}}$ is the number of load shedding stages in the UFLS scheme, $K$ is the total number of timesteps, $N_{L}$ is the number of load buses available for UFLS. The decision variable $\omega^{i}_\text{sh}$ is the frequency threshold for the $i^\text{th}$ UFLS load shedding stage and $\omega_n[k]$ is the predicted frequency at bus $n$. The constraints \eqref{eqn:opt_thresh} and \eqref{eqn:opt_binInit} ensure that $\alpha_n^i[k]$ is zero if $\omega_n[k]\geq \omega^{i}_\text{sh}$. 

\subsubsection{Time delay and deadband}
The formulation should ensure that load shed at a bus is triggered when the frequency has been below the frequency threshold ($\omega_{sh}^i$) for more than the deadband ($t_{db}$). Also, there should be a time delay of $t_{delay}$ between triggering load shed and actual actuation. Time delay and deadband are captured using the constraints defined in \eqref{eqn:Delay_and_deadband}.
\begin{subequations}\label{eqn:Delay_and_deadband}
    \begin{align}
    0\le P^i_{n} [k + 1] - P^i_{n} [k]  &\le \overline{P} \alpha^{i}_{n} [k-z]\label{eqn:opt_LoadShedStage}\\
        P^i_{n} [1]   &=0 \label{eqn:opt_LoadShedInit} \\
        \sum_{k=1}^{K}\alpha_n^i[k]&\leq K_{db} :=\frac{t_{db}}{t_s} \label{eqn:opt_DeadBand}\\
        P_\text{sh,n} [k + t_\text{delay}] &= \sum_{i=1}^{N_{UFLS}}P^i_{n}[k]\label{eqn:opt_LoadShedTotal}
    \end{align}
\end{subequations}
for $z=0,\hdots,K_{db}-1$, $K_{db}$ is the number of timesteps in the deadband, $P^i_{n}$ is the UFLS load shed trigger at bus $n$ and at the $i^{\text{th}}$ UFLS stage and $P_\text{sh}$ is the amount of load shed caused by triggering a UFLS relay and $t_s$ is the sampling time. \eqref{eqn:opt_LoadShedStage} and \eqref{eqn:opt_LoadShedInit} requires $\alpha_n^i[k]$ to be 1 for $K_{db}$ consecutive timesteps in order to trigger a change in load shed, which is only possible if the frequency was below the threshold for $K_{db}$ timesteps. \eqref{eqn:opt_DeadBand} ensures load shed at a bus can be triggered only once per UFLS stage, since a single UFLS stage does not involve multiple load shed triggers. Lastly, \eqref{eqn:opt_LoadShedTotal} ensures that the load shed at a bus is equal to the sum of the load shed at each UFLS stage, with an explicit delay to capture the time between the UFLS relay triggering and circuit breaker actuation.

\subsubsection{Load Shed Constraints}
Since over the timescale of the disturbance load shed is not reconnected back into the system. A constraint is needed to ensure that the amount of load shed at any given bus is non-decreasing with time. Also, the amount of load shed a given bus cannot exceed the amount of load available at the bus, which implies that for all times $k$ and nodes $n\in\mathcal{B}_L$, we must satisfy:
\begin{subequations}
\label{eqn:Load_Shed_Constraints}
 \begin{align}
P_{\text{sh},n}[k]&\le P^{\text{max}}_{\text{sh,}n} 
\label{eqn:PL_def} \\
0 &\le P_{sh,n}[k+1]-P_{sh,n}[k].     
\label{eqn:Monotonicity const}
\end{align}   
\end{subequations}

\subsubsection{Frequency Thresholds Constraints}
Bus frequency fluctuates typically as part of regular power system operation; therefore, UFLS setpoints set close to nominal frequency might trigger load shedding under normal conditions. Furthermore, a minimum gap between two consecutive frequency thresholds is needed to avoid overlap. The following constraints ensure a maximum frequency threshold and minimum frequency threshold gap:
\begin{subequations}\label{eqn:FreqThres_Const}
\begin{align}
        \omega^{i}_\text{sh} &\leq \omega^{\text{max}}_\text{sh}\\
        \omega^{i}_\text{sh}-\omega^{i+1}_\text{sh}  &\geq \Delta\omega_{\text{min}}.
    \end{align}
\end{subequations}

\subsection{DER Implications}
\begin{figure}[h]

   \centering
	\includegraphics[trim={2cm 0cm 10cm 0cm}, clip,width=0.8\linewidth]{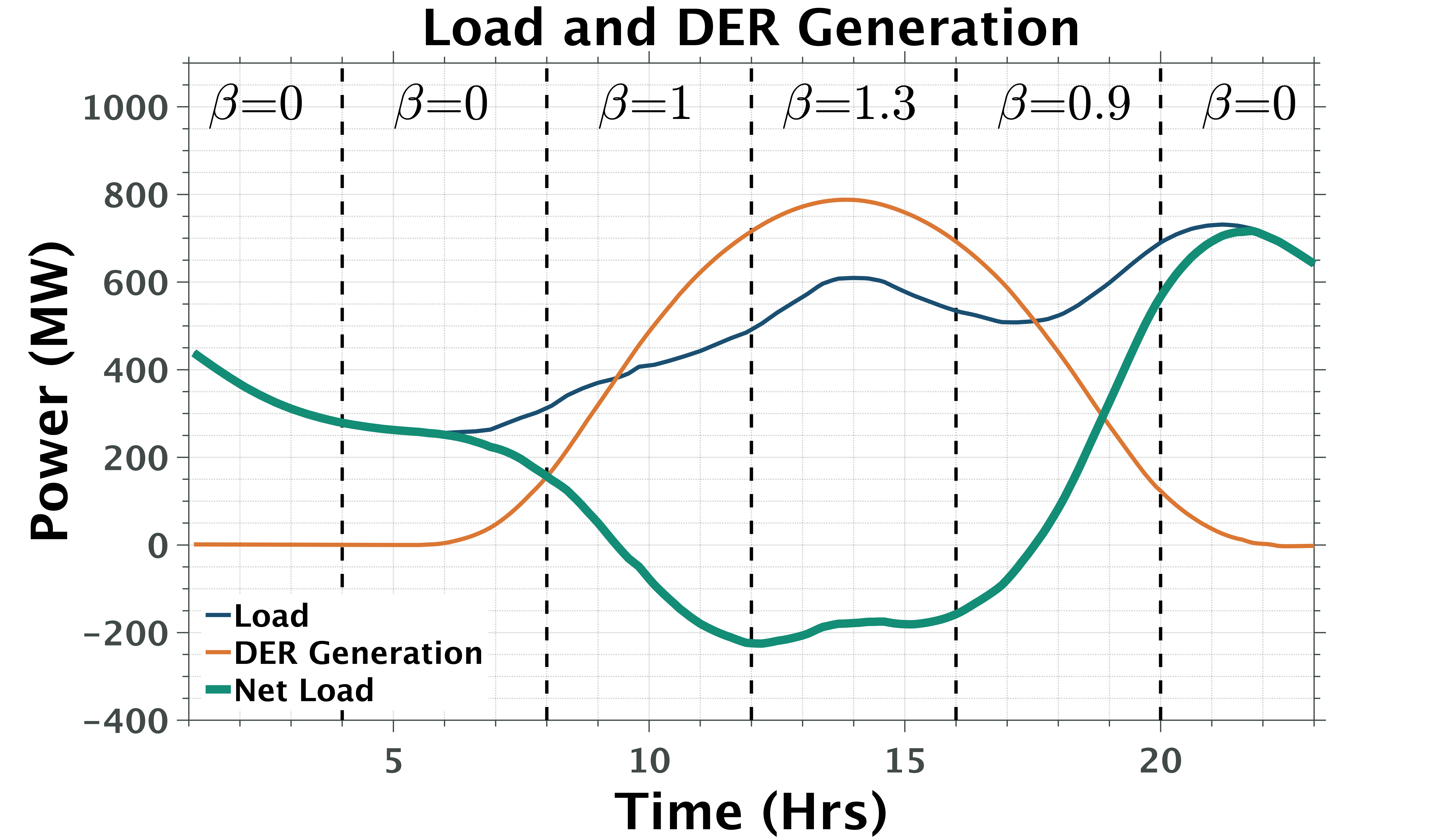}

 	\caption{Load and DER Generataion Profile and corresponding $\beta$'s}
 	\label{fig:Beta_Illustration}       
 \end{figure}

  With the prevalence of DERs such as residential PV, both load and generation are present downstream of a UFLS relay. Thus, when UFLS trips, it is the net load (i.e., the difference between load and generation) that trips. During peak PV generation and light load times, DER generation can exceed demand resulting in what is known as `back-feeding.' Shedding load at that particular bus would result in a net decrease in generation rather than a decrease in load, bringing down the frequency even further. Therefore, the presence of DERs and their relative power generation compared to the load must be considered while designing a UFLS scheme for the modern bidirectional power grid. Given a forecast of DER power generation ($\hat{P}_{DER}$) and load ($\hat{P}_{Load}$) at different load bus locations, the parameter $\beta_n$ is defined as the relative proportion of DER generation to load at a load bus and is computed as
\begin{equation}\label{eqn:Beta_Def}
    \beta_n :=\frac{\hat{P}_{DER,n}}{\hat{P}_{Load,n}},
\end{equation}
where $\hat{P}_{DER,n}$ is the forecasted DER power generation at bus $n$ and $\hat{P}_{Load,n}$ is the expected load power available at bus $n$. Note, $\beta_n>1$ indicates that there is back-feeding at bus~$n$. The introduction of the $\beta$ values allows for the distinction between the net decrease in load caused by tripping a UFLS relay and the load that was disconnected due to tripping the UFLS relay. The predicted value of $\beta$ can be computed for several time intervals and hence can be used in day-ahead planning to optimize the UFLS setpoints. Fig~\ref{fig:Beta_Illustration} illustrates how $\beta$ values would evolve over the course of a day.

Using the $\beta$ values, DER power generation can be incorporated into the optimization problem formulation as follows:
\begin{subequations}\label{eqn:DER_Shedding}
\begin{align}
    P_\text{UFLS}[k]&=P_{sh}[k]-P_{DER}[k],  \quad \forall k\\
    P_{DER,n}[k]&=\beta_n P_{sh,n}[k]\label{eqn:DER_last}
\end{align}
\end{subequations}

\noindent where, $P_\text{UFLS}[k]$ is the net load shed at a bus which is the difference between the load shed and the corresponding DER generation that was also lost.



Lastly, the bus frequencies and angles should satisfy the system model equation in \eqref{eqn:DiscretizedEqn}. Thus, the overall optimization problem becomes:
\begin{subequations}
    \label{eqn:Overall_Opt_Form}
\begin{align}
& \underset{\omega^{i}_\text{sh},P^n_\text{sh}}{\text{min}}
& & \gamma_1\sum_{n=1}^{N} ||\omega^n ||_{\infty} + \gamma_{2}\sum_{n\in \mathcal{B}_L}P^n_{\text{sh}}[K] + \gamma_3\sum_{n=1}^{N} |\omega^n[K]| \\
& \text{s. t.}
& & \eqref{eqn:DiscretizedEqn}, \eqref{eqn:Freq_Thres}-\eqref{eqn:DER_Shedding}.
\end{align}
\end{subequations}

\section{Selection Methodology}\label{Sec_Methodology}

\begin{figure}[thb]
    \centering
	\includegraphics[trim={0cm 0cm 0cm 0cm}, clip,width=0.8\linewidth]{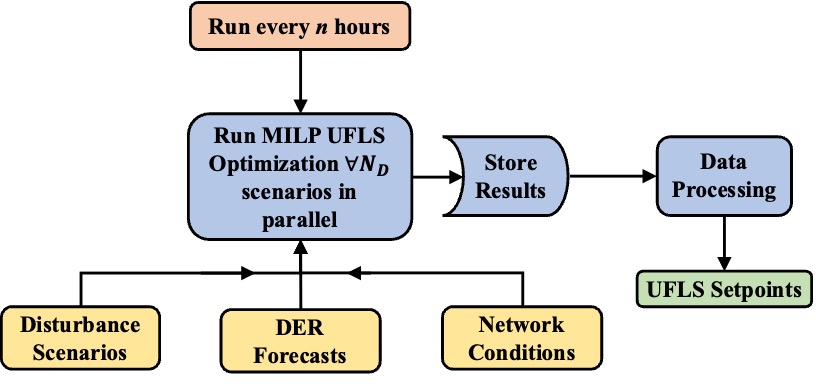}
 	\caption{Flow chart for generating optimal UFLS parameters for each scenario.}
 	 \label{FIG:FlowChart}
 \end{figure}

The optimization problem formulation in \eqref{eqn:Overall_Opt_Form} is used to find the optimal location and amounts of load-shed at each stage of UFLS. It also finds the optimal frequency thresholds for a given disturbance. According to NERC reliability standard PRC-006, a 25\% imbalance between total system load and generation is used to benchmark UFLS settings. PRC-006 lays out how long a bus frequency can be at different frequency levels, with a requirement on steady-state frequency to be between 59.3 Hz and 60.7 Hz. For the time horizon considered (20s), these requirements are met by constraining the frequency to be above 58.5 Hz and that a steady state frequency is between 59.3 Hz and 60.7 Hz. Moreover, PRC-006 mentions that the load shed for any of the UFLS stages across the entire system should not be more than 7.5\% of the total load for systems with a peak load of $\geq100$MW. With that, we seek to optimize the load shed location, amount of load shed per stage, and frequency thresholds within the stated guidelines. A 25\% load-generation imbalance, i.e. generation is lost such that there is a 25\% difference between the generation and load. This may involve losing multiple generators that add up to a 25\% imbalance throughout the system. Thus, several possible 25\% imbalance disturbance scenarios exist. In practical situations, due to the unpredictability of disturbance locations, pinpointing the exact locations of generation loss is not possible. Consequently, a set of $N_D$ possible disturbance scenarios are identified and used to benchmark UFLS settings.  However, formulating the optimization problem to include all the $N_D$ possible disturbance scenarios would increase the number of binary variables by a factor of $N_D$, increasing the complexity and computation burden significantly. Therefore, we optimize each disturbance scenario separately and obtain $N_D$ optimal UFLS settings. Subsequently, we find the mean, maximum, and minimum across all UFLS settings for each stage to make up three candidate UFLS settings. Although, the UFLS setpoints obtained directly from the MILP formulation yield the best performance for a specific disturbance scenario, practically, the disturbance scenario is unknown, making the direct use of the optimal UFLS setpoints impractical. Therefore, the minimum, maximum and mean of all of the MILP optimal UFLS setpoints are considered and their performance is compared to the ideal yet impractical optimal MILP UFLS setpoints. Fig~\ref{FIG:FlowChart} shows the process of obtaining the $N_D$ different optimal UFLS settings used to find the mean, minimum and maximum UFLS settings. In a practical setting, the optimization for each disturbance scenario can be done in parallel. This procedure would be repeated for different time intervals throughout the day and used to optimize UFLS setting in day-ahead planning.

We have identified six disturbance scenarios in the IEEE-39 bus case where the cumulative loss of generation results in approximately 25\% imbalance between load and generation. Using these scenarios

 Note, it was observed that the maximum and minimum UFLS setpoints discussed earlier yielded poor performance and have a tendency to overshed and undershed respectively. Hence, the mean UFLS setpoints will be considered and benchmarked against the impractical optimal MILP setpoints of the corresponding disturbance scenario to assess the loss in optimality associated with the averaging.

We compare the system performance for each disturbance scenario with i) the optimal UFLS settings for that disturbance ii) the mean UFLS settings, and iii) the conventional UFLS settings laid out in PRC-006. The three sets of UFLS setpoints are summarized as follows:
\begin{enumerate}
    \item \textbf{Average of Optimal Setpoints (AOS)}: Obtained by taking the mean of the optimal UFLS setpoints across all disturbance scenarios.
    \item \textbf{Ideal Case Setpoints}: The optimal UFLS setpoints obtained from the proposed MILP optimization formulation for a given disturbance. These setpoints are the \emph{unrealistic/ideal case scenario}, requiring omniscient prior knowledge of the disturbance size and location, which is impossible in a practical setting. They represent the best possible UFLS scheme performance and are used to asses the loss in optimality associated with the MOS setpoints. 

    \item \textbf{Conventional}: The UFLS setpoints as laid out in PRC-006 by NERC.
\end{enumerate}

\section{Simulation and Results } \label{Sec_ResultsDiscussion }
The proposed optimization-based UFLS scheme is evaluated on the IEEE 39 bus system. Six disturbance scenarios were identified, each involving a loss of multiple generators, causing approximately a  25\% imbalance between load and generation. The optimal UFLS setpoints for each disturbance scenario are obtained and then used to obtain the AOS UFLS setpoints by averaging the setpoints across all disturbance scenarios. The performance of the AOS setpoints is compared to Ideal Case setpoints and the conventional UFLS setpoints laid out by NERC in PRC-006. The following parameters are used for our analyses: $\Delta\omega_\text{min} = 0.1 \text{Hz},$  $\omega^\text{max}_\text{sh} = 59.5 \text{Hz}$, a dead-band of $300 \text{ms}$, and a $100 \text{ms}$ delay between relay trigger and circuit breaker actuation. Each load shedding stage is limited to 7.5\% of the total load, and the load shed at each bus is limited to the amount of load available at the bus. 

Furthermore, four cases are identified to test the performance of the proposed scheme under multiple scenarios. The first case is the base case and assumes nominal operating conditions with generation coming solely from synchronous generation and no DER generation present at the load buses. The second case considers DER generation is present at the load buses. To simulate such a scenario $\beta$ values were randomly generated for the 20 load buses in the 39-bus system with four random buses having $\beta>1$, indicating back-feed. Note, the introduction of DER's reduces the net-load in the system; therefore, the generation is reduced in proportion to the reduction in net system load. The third case considers DER generation at load buses and 50\% reduction in inertia. The reduction in inertia is meant to capture the scenario where inverter based resources (IBR) displace synchronous generation resulting in a system with lower inertia and hence more sensitive to power imbalances. In the previous three cases both the frequency thresholds and the amount of shed were modified to yield the Ideal case and, in turn, the AOS setpoints. This raises the question of whether the proposed scheme's improved performance is due to its larger degree of freedom when selecting UFLS setpoints compared to the conventional UFLS scheme. Consequently, a fourth case is considered where the frequency thresholds are fixed to be equal to that of the conventional UFLS scheme. The only decision variables in this case are the location and amount of load to shed for a better comparison.

\begin{figure}[t]

   \centering
	\includegraphics[trim={0cm 0cm 10cm 0cm}, clip,width=0.8\linewidth]{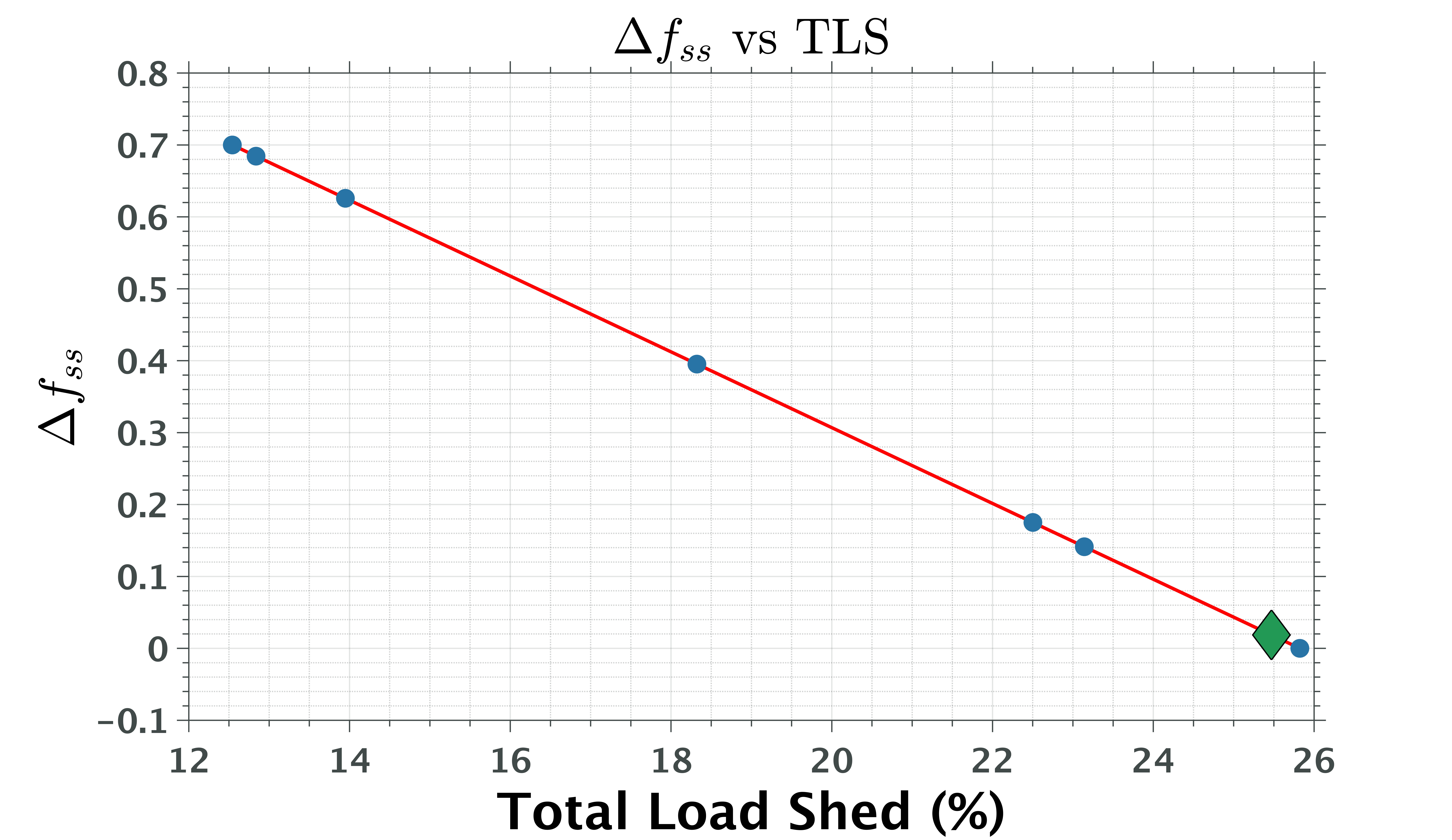}

 	\caption{Trade-off curve for objectives TLS and $\Delta f_{\text{ss}}$ }
 	\label{fig:Pareto}       
 \end{figure}

The following metrics were used to assess the performance of each of the UFLS schemes in arresting the frequency decline during a disturbance:
\begin{enumerate}
\item \emph{Frequency Nadir}: Bus frequencies need to be within acceptable limits to ensure power system stability. The frequency nadir is the minimum system-wide bus frequency reached following a disturbance.
\item \emph{Total Amount of Load Shed (TLS)}: The total amount of load shed across all buses at the end of the time horizon (20 s) considered for analyzing the disturbance.
\item \emph{Steady State Frequency Deviation ($\Delta f_{\text{ss}}$)}: The frequency deviation from 60 Hz after the system has reached steady state.
\end{enumerate}
The stated metrics correspond to the individual terms in the multi-objective function defined in~\eqref{eqn:Objective}. In choosing the relative weights $\gamma_1,\gamma_2,\gamma_3$ in objective function~\eqref{eqn:Objective}, the effect of the choice of $\gamma_1,\gamma_2 \text{ and }\gamma_3$ on the three performance metrics listed. The three terms in~\eqref{eqn:Objective} were first normalized by their respective worst-case values and then the values of $\gamma_1,\gamma_2 \text{ and }\gamma_3$ were varied within $[0,1]$. It was observed that the frequency nadir was not significantly affected by the value of $\gamma_1$ ($<50$ mHz change), indicating that the frequency nadir is more dictated by how fast the UFLS scheme can react to the frequency decline which is mainly limited by the time delay and deadband. The relationship between TLS and $\Delta f_{\text{ss}}$ was also studied as the value of $\gamma_2$ changed in relation to $\gamma_3$. A linear relationship between TLS and $\Delta f_{\text{ss}}$ is observed and illustrated in Fig.\ref{fig:Pareto} which shows (TLS, $\Delta f_{\text{ss}}$) for varying $\gamma$ values. It is observed that the TLS and $\Delta f_{\text{ss}}$ exhibit discrete changes as the values of $\gamma$'s are varied which is not surprising given the discrete switching nature of the system.  In this work, $\gamma_1,\gamma_2 \text{ and }\gamma_3$ are set to be 1, 0.2, and 1 respectively (green diamond in Fig~\ref{fig:Pareto}) prioritizing steady-state frequency and frequency nadir over TLS.

\subsection{Case I: Base Case}
\begin{figure}[t]

   \centering
	\includegraphics[trim={0cm 0cm 10cm 0cm}, clip,width=0.9\linewidth]{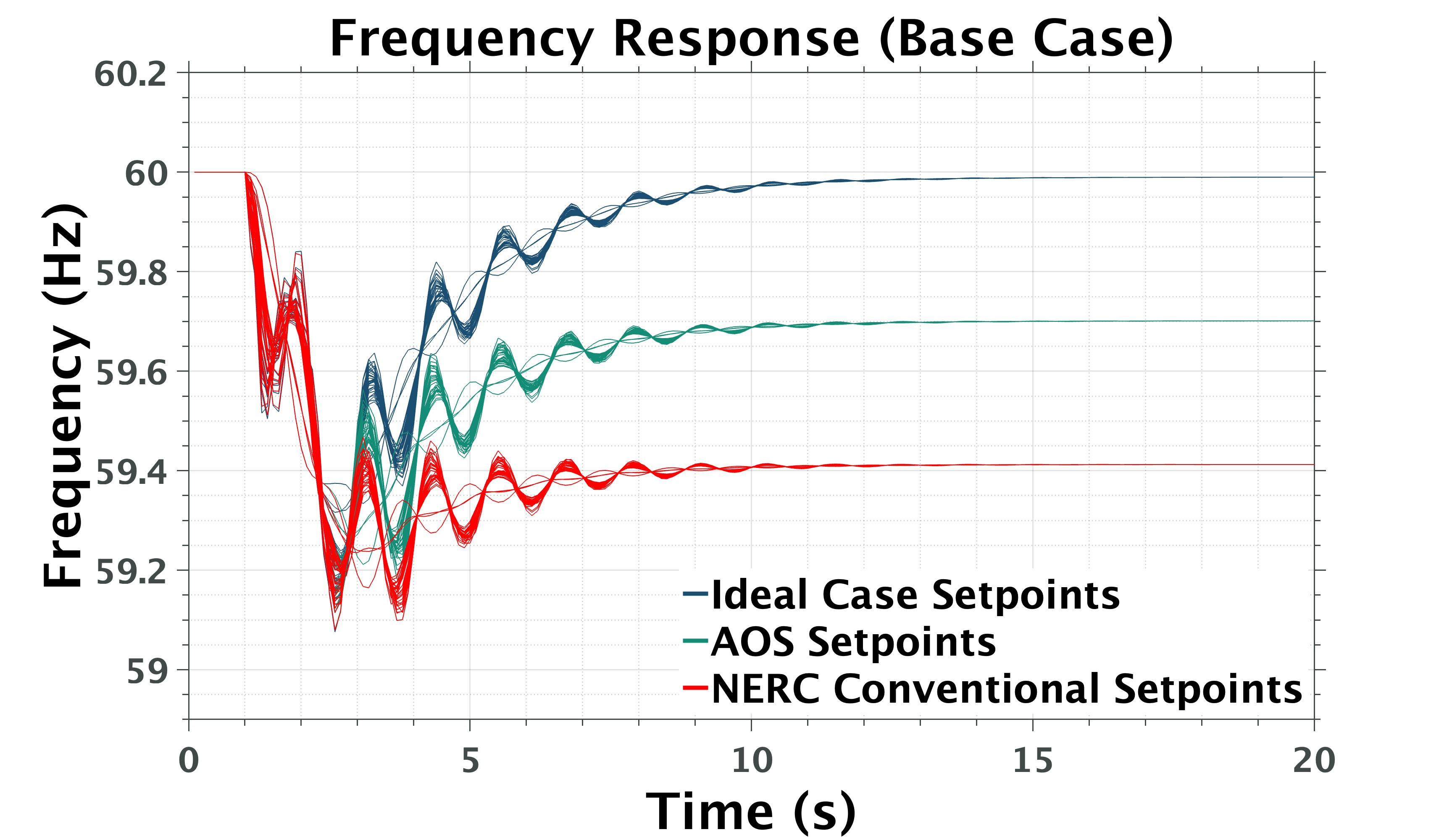}

 	\caption{Frequency Response IEEE 39 Bus}
 	\label{fig:NoDER_High}       
 \end{figure}

In the base case scenario, the nominal parameter values of the IEEE 39 bus system are utilized without any DER generation or changes in inertia. For the six disturbance scenarios, the frequency nadir, total load shed and the steady state frequency deviation are obtained. Table~\ref{tab:No_DER_High} summarizes the worst case frequency nadir, total load shed and the steady state frequency across all six disturbance scenarios. Note the worst case metric values may not correspond to the same disturbance scenario. Fig~\ref{fig:NoDER_High} shows the bus frequencies following a disturbance with each of the UFLS setpoints (Ideal, AOS and Conventional) implemented.

\begin{table}[h]
\centering
\caption{UFLS scheme performance metrics (Base Case)}
\label{tab:No_DER_High}
\begin{tabular}{@{}llll@{}}
\toprule
\multicolumn{1}{c}{\textbf{Method}} & \multicolumn{1}{c}{\textbf{Nadir (Hz)}} & \multicolumn{1}{c}{\textbf{TLS (\%)}} & \multicolumn{1}{c}{\textbf{$\Delta f_{\text{ss}}$ (Hz)}} \\ \midrule
Ideal Setpoints & 59.03 & 25.96 & 0.01 \\
AOS Setpoints & 59.02 & 25.32 & 0.30 \\
NERC Conventional Setpoints & 59.02 & 15.00 & 0.59 \\ \bottomrule
\end{tabular}
\end{table}

All three UFLS setpoints produce similar frequency nadir values. The conventional UFLS setpoints are observed to under-shed load, resulting in the poorest performance in terms of the steady state frequency. The AOS UFLS setpoints result in an improved steady-state frequency compared to the conventional UFLS scheme. 

\subsection{Case II: DERs at Load Buses}
\begin{figure}[t]

   \centering
	\includegraphics[trim={0cm 0cm 10cm 0cm}, clip,width=0.9\linewidth]{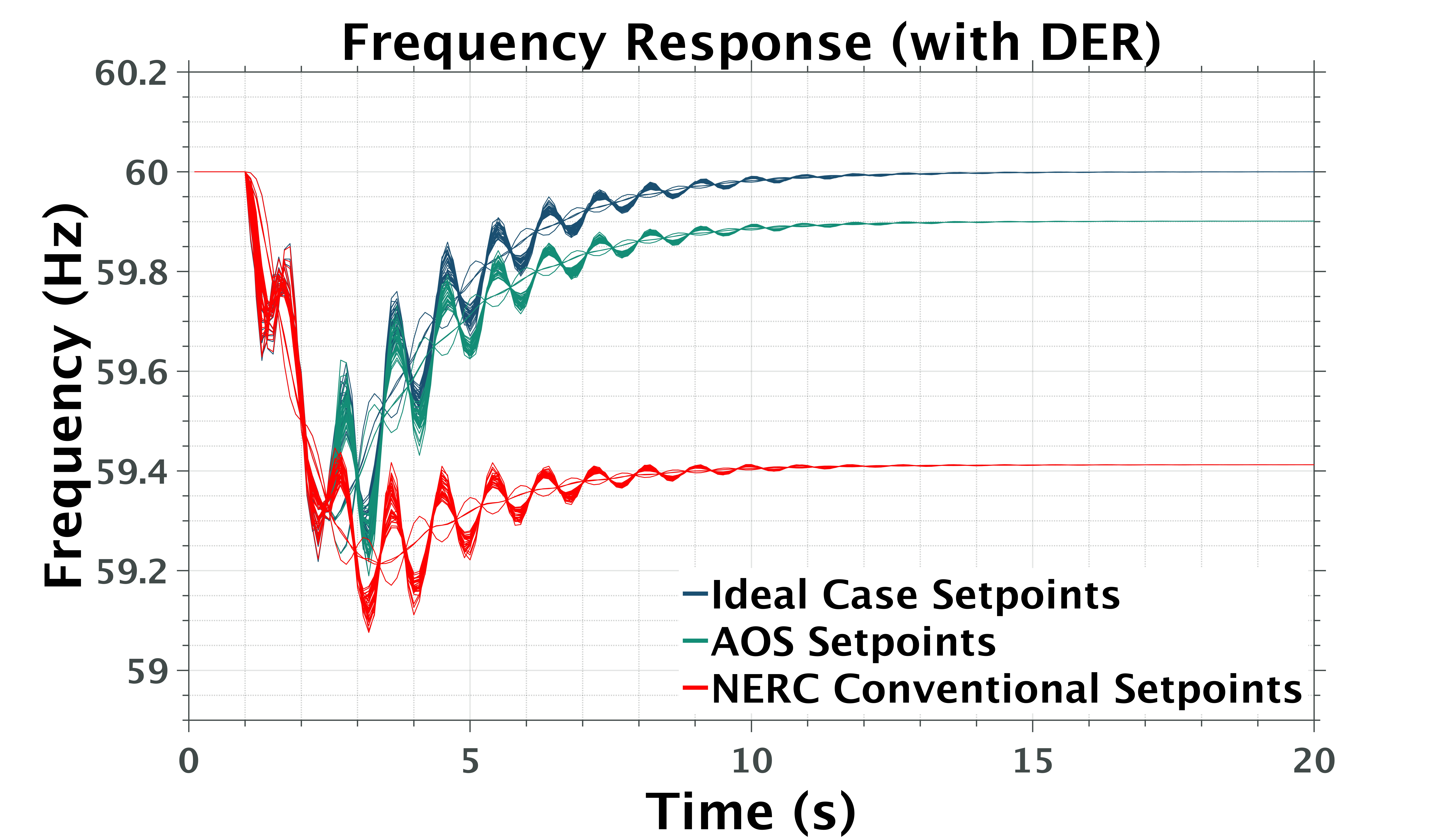}

 	\caption{Frequency Response IEEE 39 Bus with DER Generation at Load Buses}
 	\label{fig:DER_High}       
 \end{figure}

In this case, we introduced distributed energy resources (DERs) at the load buses. Three buses are randomly selected to have $\beta \in [1.1,1.3]$ to simulate back-feeding at those buses. The $\beta$ values for the rest of the buses are selected such that $\beta \in [0,0.15]$, meaning anywhere between 0-15\% of the load at a load bus is supplied by DER generation at that bus. With DERs now introduced at the load buses, there is a heterogeneity in shedding load. Two relays with the same amount of net power demand behind them could have vastly different amounts of load connected with the introduction of DER generation behind the UFLS relays.
Table~\ref{tab:DER_High} summarizes the worst-case frequency nadir, total load shed and the steady state frequency across all six disturbance scenarios. Fig~\ref{fig:DER_High} shows the bus frequencies following a disturbance with each of the UFLS setpoints (Ideal, AOS and Conventional) implemented.

\begin{table}[h]
\centering
\caption{UFLS scheme performance metrics (with DERs)}
\label{tab:DER_High}
\begin{tabular}{@{}llll@{}}
\toprule
\multicolumn{1}{c}{\textbf{Method}} & \multicolumn{1}{c}{\textbf{Nadir (Hz)}} & \multicolumn{1}{c}{\textbf{TLS (\%)}} & \multicolumn{1}{c}{\textbf{$\Delta f_{\text{ss}}$ (Hz)}} \\ \midrule
Ideal Setpoints & 59.12 & 26.98 & 0.01 \\
AOS Setpoints & 59.11 & 24.99 & 0.10 \\
NERC Conventional Setpoints & 59.07 & 26.54 & 0.59 \\ \bottomrule
\end{tabular}
\end{table}

\begin{figure}[thb]
    \centering
	\includegraphics[trim={0cm 0cm 0cm 0cm}, clip,width=0.8\linewidth]{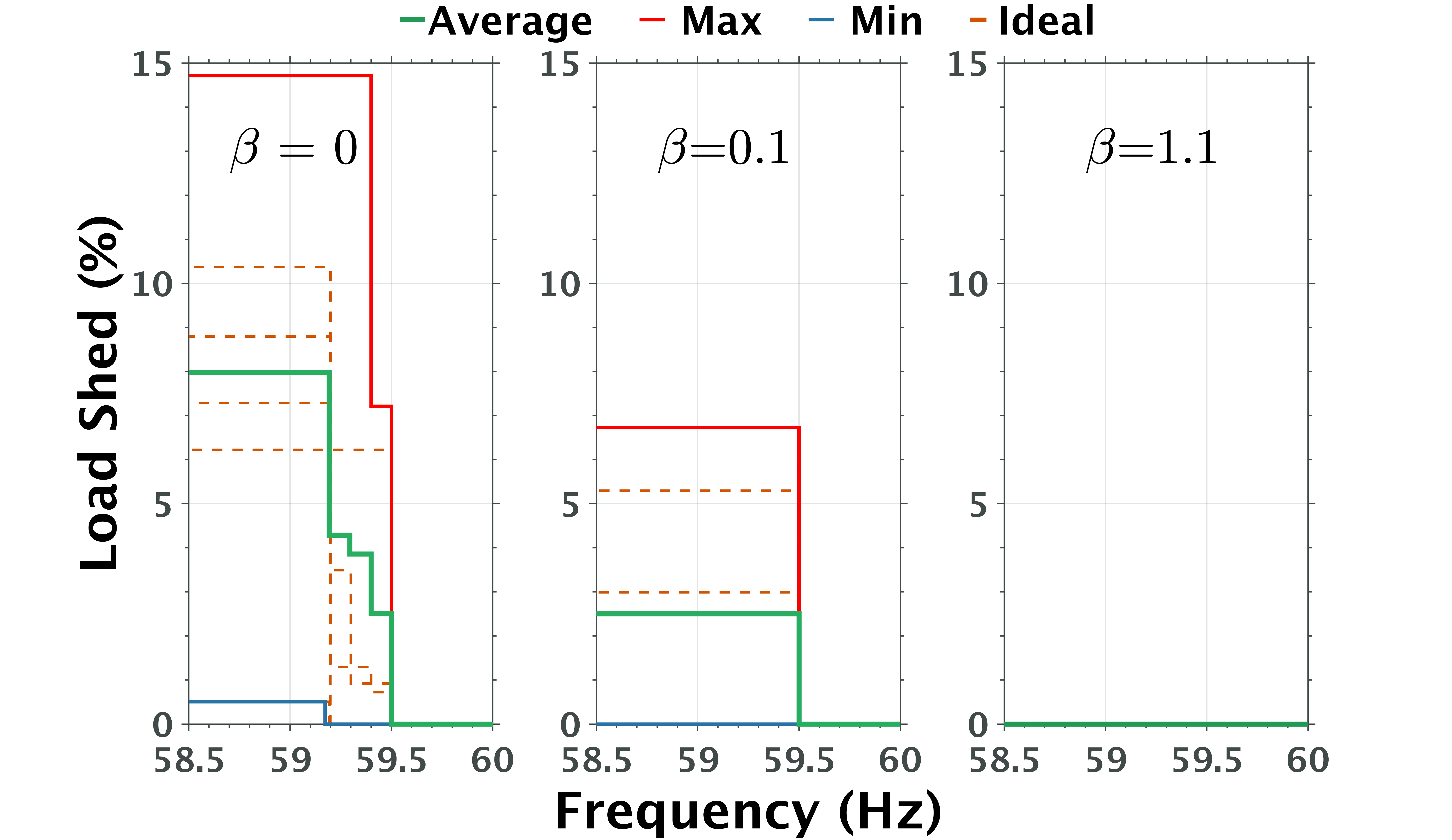}
 	\caption{UFLS setpoints for different DER penetration scenarios.}
 	 \label{FIG:Setpoints}
 \end{figure}

Table~\ref{tab:DER_High} shows that despite shedding a similar amounts of load, the AOS setpoints and ideal setpoints perform better in terms of frequency nadir and steady state frequency. 
 The conventional UFLS scheme operates statically, not adjusting to DER generation at various buses. In contrast,  the MILP-based setpoints (AOS and Ideal) select buses optimal for UFLS, minimizing load shedding for the same power demand reduction. Additionally, unlike the conventional method, the AOS and Ideal setpoints avoid back-feeding buses entirely.  This is illustrated in Fig~\ref{FIG:Setpoints} which shows the UFLS load shed setpoints for varying levels of DER penetration or, i.e., for varying values of $\beta$. Note, that in the case when $\beta>1$, indicating back-feeding, no load is shed at that bus.

\subsection{Case III: DERs at Load Buses and Low Inertia}
\begin{figure}[t]

   \centering
	\includegraphics[trim={0cm 0cm 10cm 0cm}, clip,width=0.9\linewidth]{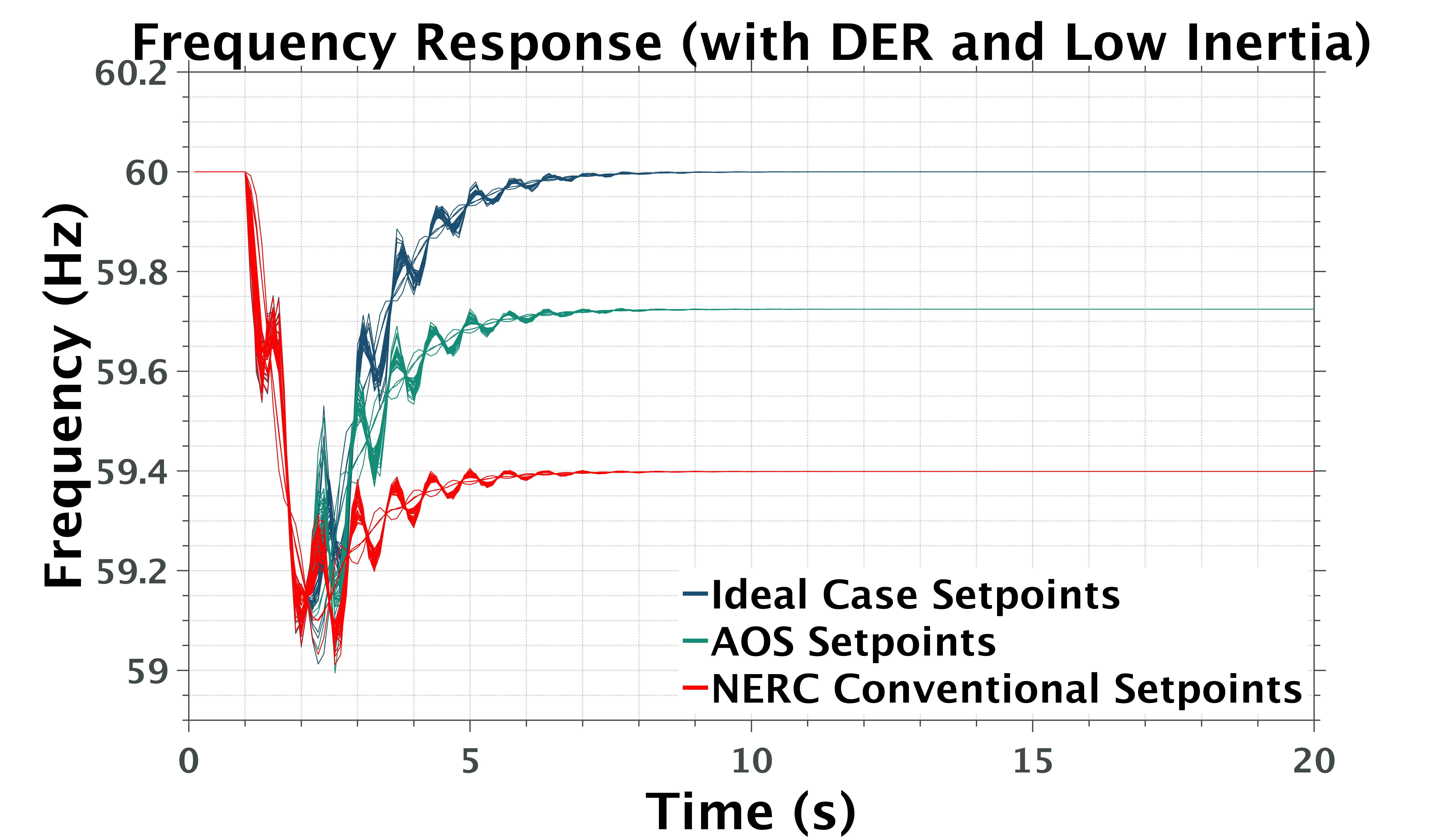}

 	\caption{Frequency Response IEEE 39 Bus with DER Generation at Load Buses and Low Inertia}
 	\label{fig:DER_Low}       
 \end{figure}

In this case, the inertia at the generator buses is reduced to half of the nominal value to simulate the effect of IBR's displacing synchronous generation resulting in a system with lower inertia. Table~\ref{tab:DER_Low} summarizes the worst-case frequency nadir, total load shed and the steady state frequency across all six disturbance scenarios. Fig~\ref{fig:DER_Low} shows the bus frequencies following one of the disturbances with each of the UFLS setpoints (Ideal, AOS and Conventional) implemented.

\begin{table}[h]
\centering
\caption{UFLS scheme performance metrics (with DERs and Low Inertia)}
\label{tab:DER_Low}
\begin{tabular}{@{}llll@{}}
\toprule
\multicolumn{1}{c}{\textbf{Method}} & \multicolumn{1}{c}{\textbf{Nadir (Hz)}} & \multicolumn{1}{c}{\textbf{TLS (\%)}} & \multicolumn{1}{c}{\textbf{$\Delta f_{\text{ss}}$ (Hz)}} \\ \midrule
Ideal Setpoints & 58.95 & 26.80 & 0.01 \\
AOS Setpoints & 58.91 & 21.46 & 0.28 \\
NERC Conventional & 58.91 & 33.88 & 0.59 \\ \bottomrule
\end{tabular}
\end{table}

 Similar to the previous case, the MILP based methods were able to achieve a better worst case nadir and steady state frequency deviation despite causing less amount of load shed than the conventional UFLS scheme. 

\subsection{Case IV: Fixed UFLS Frequency Thresholds}
\begin{figure}[t]
   \centering
	\includegraphics[trim={0cm 0cm 10cm 0cm}, clip,width=0.9\linewidth]{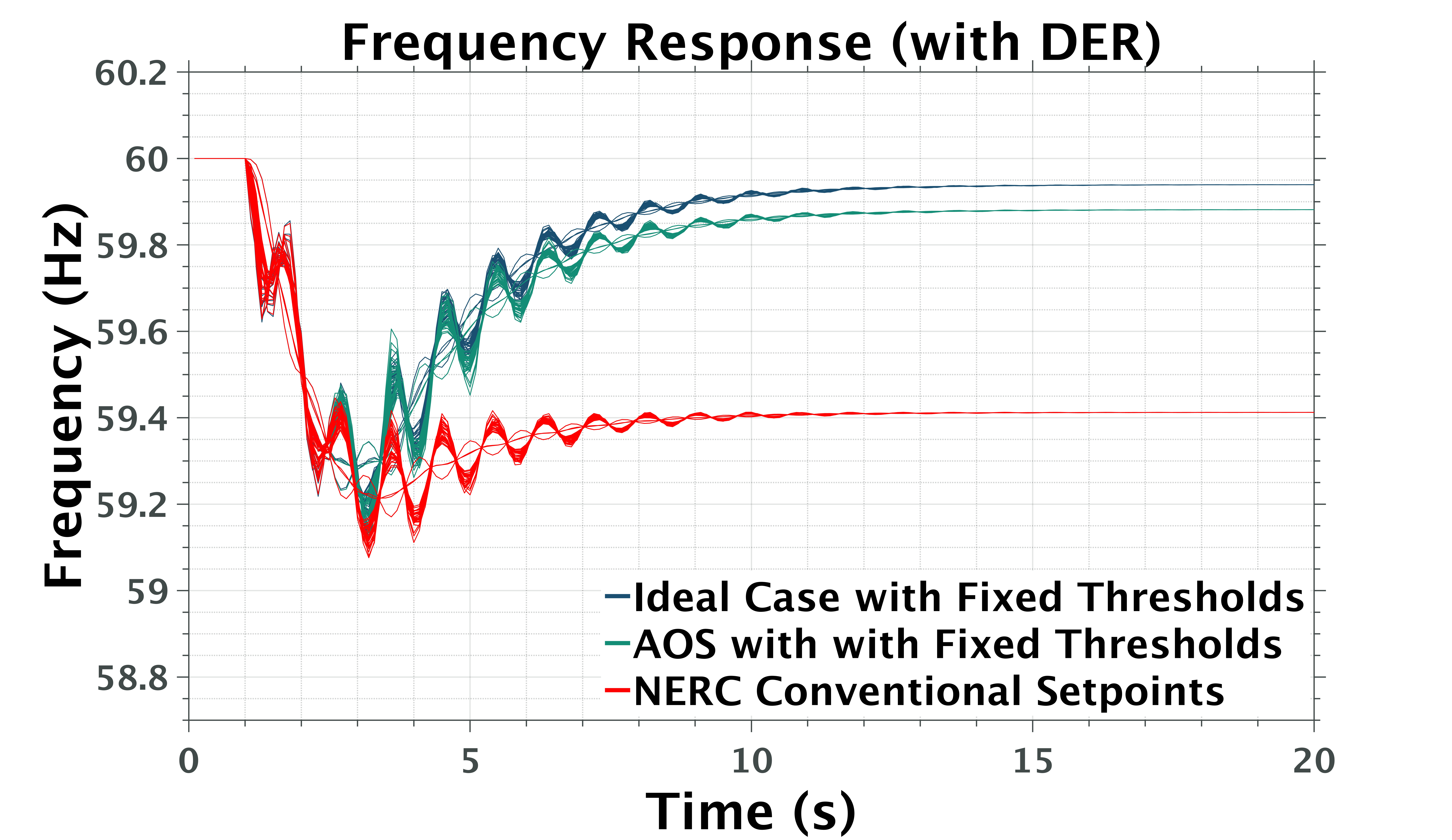}
 	\caption{Frequency Response IEEE 39 Bus with Fixed  Frequency Thresholds (with DERs) }
 	\label{fig:Fixed_Threshold_DER}       
 \end{figure}

The previous cases improved UFLS scheme performance by adapting both the frequency thresholds and the load shed amounts. For a better comparison with the conventional UFLS scheme the UFLS frequency thresholds are constrained to be equal to that of the conventional UFLS scheme for a better comparison. It was observed that in the case without DER generation present at the demand side, all three UFLS schemes (Ideal, AOS and Conventional) yield similar performance. However, with DERs introduced, the MILP-based methods were able to leverage the spatial DER generation awareness to optimally shed load at buses with less DER generation, resulting in a larger reduction in power demand for the same amount of load shed. Fig.~\ref{fig:Fixed_Threshold_DER} shows the bus frequencies after the disturbance while using the three UFLS schemes (Ideal, AOS and Conventional) respectively. Table~\ref{tab:FixedW_DER_High} summarizes the worst case frequency nadir, total load shed and the steady state frequency across all six disturbance scenarios.

\begin{table}[]
\centering
\caption{UFLS scheme performance metrics with fixed frequency Thresholds (with DERs)}
\label{tab:FixedW_DER_High}
\begin{tabular}{@{}llll@{}}
\toprule
\multicolumn{1}{c}{\textbf{Method}} & \multicolumn{1}{c}{\textbf{Nadir (Hz)}} & \multicolumn{1}{c}{\textbf{TLS (\%)}} & \multicolumn{1}{c}{\textbf{$\Delta f_{\text{ss}}$ (Hz)}} \\ \midrule
Ideal Setpoints & 59.12 & 26.98 & 0.01 \\
AOS Setpoints & 59.12 & 24.99 & 0.27 \\
NERC Conventional Setpoints & 59.09 & 26.54 & 0.59 \\ \bottomrule
\end{tabular}
\end{table}

\section{Conclusion} \label{Sec_conclusion }

A mixed-integer linear programming (MILP) formulation was presented to obtain optimal UFLS setpoints in bulk power grids that adapt with changing network conditions. The MILP formulation utilizes a linear power system model that was validated against more accurate non-linear power system dynamics simulators (PSAT). The proposed MILP formulation halves the number of binary variables necessary while still incorporating 1) conditions when the frequency is predicted to be below a threshold and 2) triggering the load shedding action.  The formulation also accounts for forecasted DER generation at each load bus to optimally embed UFLS participation factors into the resulting UFLS parameters. This was shown to yield improved UFLS scheme performance with equivalent or even less amount of load shed compared to the conventional UFLS scheme. The formulation also mitigates the growing concern of triggering a UFLS relay at a load bus that is back-feeding due to DER injections, which would worsen the frequency response and could cause wide-scale blackouts.

Despite the MILP formulation being more efficient, it is still non-convex (due to binary variables), which results in a computationally complex approach. In addition, a solution is only optimal with respect to its specific, pre-determined disturbance (scenario). Thus, the MILP methodology must compute optimal solutions across all salient scenarios.
{From this database of solutions, we propose a methodology that employs the mean UFLS settings across the identified disturbance scenarios. Although the averaging of the UFLS setpoints results in sub-optimal solution, the averaged dynamic setpoints are observed to out-perform the conventional UFLS setpoints, even when the frequency thresholds are constrained to be the same as that of the conventional UFLS scheme.

Future work will focus on scaling up the formulation by leveraging nonlinear programming formulations such as~\cite{AmritNLP} that offer scale and can serve to provide the MILP formulation with a near-optimal, integer-feasible warm-start. Furthermore, we are interested to investigate and improve robustness of the approach by incorporating the uncertainty inherent to DER injections (i.e., account for DER forecast errors).

\appendix \label{Sec_appendix }

The system ODE's can be expressed in the following form:
\begin{equation}
    \dot{x}_G=Ax+BP_{\text{UFLS}}+G\Delta
\end{equation}
where $x^\top:=\text{col}\{\theta^\top, \omega^\top, x_\text{gov}^\top\}$ and  $x_G^\top:=\text{col}\{\theta_G^\top, \omega_G^\top, x_\text{gov}^\top\}$

The system matrix A is given by
\begin{align}\label{eqn:A_Matrix}
            A:=\begin{bmatrix}
                \mathbf{0} & A_{12} & \mathbf{0} \\ 
                A_{21} & A_{22} & A_{23} \\ 
                \mathbf{0} & A_{32} & A_{33}
            \end{bmatrix},
\end{align}
where block matrices $A_{ij}$ are defined as
\begin{align*}
    A_{12}:&= \omega_\text{base}\Pi_G \qquad 
    &A_{21}:= -M^{-1}\Pi_G \\
    A_{22}:&= -M^{-1}D\Pi_G \qquad 
    &A_{23}:= -M^{-1}K_\text{gov} \\
    A_{32}:&= T^{-1}_\text{gov}  \Pi_G \qquad 
    &A_{33}:= -T^{-1}_\text{gov}.
\end{align*}

$\Pi_G$ is a selection matrix and is defined such that it satisfies $
\omega_G=\Pi_G \omega$. The matrices B and G can be expressed as
\begin{align}\label{eqn:B_Matrix}
            B:=\begin{bmatrix}
            0 \\
            -M^{-1}\Pi_G  \\
            0
            \end{bmatrix}
            \quad 
            \text{ and }
            \quad 
             G:=\begin{bmatrix}
            0 \\
            -M^{-1}\Pi_G  \\
            0
            \end{bmatrix}.
\end{align}


\bibliographystyle{IEEEtran}
\bibliography{sample.bib}

\end{document}